\documentclass[aps,amsmath,amssymb,twocolumn,superscriptaddress,
prl,showkeys,array]{revtex4}

\usepackage[dvips]{graphicx}

\begin{document}
\draft


\title{Two-state behaviour of Kondo trimers}
\author{M. E. Torio}
\address{Instituto de F\'{i}sica Rosario, CONICET-UNR,\\
Bv 27 de Febrero 210 bis, (2000) Rosario, Argentina \\
} 
\author{K. Hallberg}
\address{Centro At\'{o}mico Bariloche and Instituto Balseiro, CNEA\\ 8400
Bariloche, Argentina} 
\author{C. R. Proetto}
\address{Centro At\'{o}mico Bariloche and Instituto Balseiro, CNEA\\ 8400
Bariloche, Argentina} 


\begin{abstract} 

The electronic properties and spectroscopic features of a magnetic trimer
with a Kondo-like coupling to a non-magnetic metallic substrate are analyzed
at zero temperature. The substrate density of states is depressed in the
trimer neighbourhood, being exactly zero at the substrate chemical
potential. The size of the resonance strongly depends on the magnetic state
of the trimer, and exhibits a two-state behavior. 
The geometrical dependence of these results agree qualitatively with recent 
experiments and could be reproduced in a triangular quantum dot arrangement.

\end{abstract}

\maketitle


\narrowtext

The combination of the Scanning Tunneling Microscope (STM), which can be
used as a ``tool'' to build and characterize atomically-precise structures,
and the availability of high-quality clean metallic surfaces has provided a
unique opportunity to investigate the chemical and physical properties of
deposited atoms, molecules, and clusters with unprecedented resolution. It
has been natural then that much of the research concentrates on one of the
old paradigms of condensed-matter physics: the behavior of magnetic
impurities immersed in an otherwise non-magnetic matrix. Ground-breaking
experiments have been reported on the properties of magnetic monomers\cite
{li} and dimers\cite{chen}, quantum-confinement of surface two-dimensional
electron gases by a ``corral'' of STM-positioned magnetic atoms\cite
{manoharan}, and ``molecule cascade'' devices\cite{eigler}, with potential
for quantum computing. In most of these examples, it has become evident
that at low-temperatures the physics of magnetic impurities on the metallic
surface is dominated by the so-called Kondo effect. Historically, the Kondo
effect was introduced more than forty years ago to explain the resistivity
minimum for decreasing temperatures observed in metallic matrices with a
minute fraction of magnetic impurities.\cite{jun} 
Additional interest in the Kondo effect arose recently from the
semiconductor nano-oriented community, after realization\cite
{glazman,patrick} that a quantum dot with two attached conducting leads
behaves in many aspects as a magnetic atom in a metal.\cite{david,sara}

The aim of this work is the theoretical study of a magnetic trimer in
contact with a metallic substrate. The trimer geometry is very interesting,
due to the fact that it is the simplest geometrical arrangement (including
monomers and dimers) which displays magnetic frustration among the three
magnetic moments for the case of antiferromagnetic interactions\cite{ingersent}. The 
main
result from our work is that as a consequence of the interplay between the
trimer-metal Kondo interaction which tries to stabilize the
trimer in a high-spin state, and the intra-trimer antiferromagnetic
interactions which try of stabilize a low-spin trimer state, the whole
system exhibits a two-state spectroscopic behavior that depends on the geometrical 
arrangement of the triangle (whether it is an equilateral or isosceles triangle) and 
on the interaction parameters. 

\begin{figure}[tbp]
\includegraphics[width=0.45\textwidth]{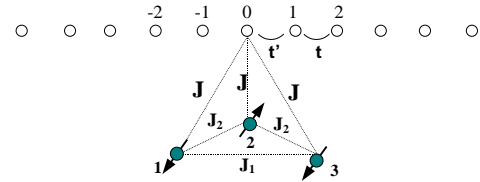}
\vspace{-0.5cm} 
\caption{Schematic diagram of the system considered: an AF spin trimer 
Kondo-coupled to a non-interacting chain} 
\end{figure}

The model employed in the calculation 
consists of two semi-infinite non-interacting tight-binding (TB) chains
connected to a central site (also non-interacting). This central site is in
turn Kondo-coupled to three magnetic impurities, the trimer.The
Hamiltonian reads: 
\begin{equation}
H=H_{c}+H_{c-t}+H_{t},
\end{equation}
where $H_{c}$ is the Hamiltonian of two semi-infinite chains, 
\begin{eqnarray}
H_{c}&=&-\sum_{j\leq 0,\sigma }t_{j-1}(c_{j\sigma }^{\dagger }c_{j-1\sigma
}+\text{h.c.})  \nonumber \\
&-&\sum_{j\geq 0,\sigma }t_{j+1}(c_{j\sigma }^{\dagger }c_{j+1\sigma }+\text{%
h.c.});  \eqnum{2a}
\end{eqnarray}
$H_{c-t}$ couples the chains to the trimer and can be written as 
\begin{equation}
H_{c-t}=\sum_{\alpha ,\sigma }t''_{\alpha}(c_{0\sigma }^{\dagger }d_{\alpha
\sigma }+\text{h.c.}),  \eqnum{2b}
\end{equation}
and $H_{t}$ is the trimer Hamiltonian 
\begin{equation}
H_{t}=\sum_{\alpha\ne \beta ,\sigma }t_{\alpha \beta } d_{\alpha \sigma
}^{\dagger }d_{\beta \sigma } +\sum_{\alpha }\left(
\varepsilon _{0}n_{\alpha }+U\text{ }n_{\alpha \uparrow }n_{\alpha
\downarrow }\right) .  \eqnum{2c}
\end{equation}
In the equations above, $t_{j}=t=1$ for $j\leq -1$, $j\geq 1,$ while $%
t_{-1}=t_{1}=t^{\prime }=1$,  $\alpha ,\beta =1,2,3$ (the trimer
sites). $c_{j\sigma }$ $(c_{j\sigma }^{\dagger })$ corresponds to the destruction
(creation) operator for one electron at the chain site $j$ with spin $\sigma $ $%
(=\uparrow ,\downarrow ),$ while $d_{\alpha \sigma }$ $(d_{\alpha \sigma
}^{\dagger })$ is a destruction (creation) operator for electrons at the
trimer sites. $n_{\alpha \sigma }=d_{\alpha \sigma }^{\dagger }d_{\alpha
\sigma },$ $n_{\alpha }=\sum_{\sigma }n_{\alpha \sigma },$ and $U>0$. If $%
t''_{\alpha}=0$ for $\alpha=1,2,3$, $H_{c-t}=0$ and the whole system becomes the sum of
two uncoupled subsystems: the tight-binding chain and the trimer. In the
Kondo regime it is feasible to simplify the Hamiltonian above by performing
a Schrieffer-Wolff (SW) canonical transformation of the second and
third terms in Eq.(1), which projects out of the Hilbert space states where
the trimer sites are either empty or double occupied. Keeping only the most
relevant terms, the result of the transformation for these terms is (see Fig. 1)

\begin{equation}
H_{c-t}+H_{t}=\sum_{\alpha }\bar{J}_{\alpha}\,({\bf s}_{0}\cdot {\bf S}_{\alpha
})+\sum_{\alpha\ne \beta }\bar{J}_{\alpha \beta }\,({\bf S}_{\alpha }{\bf \cdot S}%
_{\beta }).  \eqnum{3}
\end{equation}
The first term in Eq.(5) represents an {\it s-d }interaction among the trimer
spins and the central site of the chain; the second term is a Heisenberg 
interaction among the spins located at each of the trimer sites. For the
symmetric case $2\varepsilon _{0}+U=0$ considered in the present work, $%
\bar{J}_{\alpha}=8\left| t''_{\alpha }\right| ^{2}/U,$ $\bar{J}_{\alpha \beta }=8\left| 
t_{\alpha \beta }\right|^{2}/U.$ As $U>0,$ both $\bar{J}_{\alpha}$ and $\bar{J}_{\alpha 
\beta }$ are positive, corresponding to antiferromagnetic
interactions in Eq.(5). Unless otherwise stated, we will assume in this work
a symmetric coupling configuration: $\bar{J}_{1}=\bar{J}_{2}=\bar{J}_{3}=J,$ and $%
\bar{J}_{13}=J_{1},$ $\bar{J}_{12}=\bar{J}_{23}=J_{2}.$ Numerical calculations have been
performed for both Anderson-like (Eqs.(2-4)) and Kondo/Heisenberg-like
models (Eq.(2),(5)), but results will be shown only for the latter. 
Results considering Anderson impurities in the SW limit present a qualitatively 
similar physical behaviour.

Since the possible trimer states will play a relevant role in the following
discussions, we will write down here its eigenvectors and eigenvalues. From
Eq.(5) we have $H_{t}=J_{1}{\bf S}_{1}{\bf \cdot S}_{3}+J_{2}({\bf S}_{1}%
{\bf \cdot S}_{2}+{\bf S}_{2}{\bf \cdot S}_{3}).$ If $J_{1}=J_{2}$ the
trimer is in the {\it equilateral} (E) configuration, whereas if $J_{1}\neq J_{2}$
the trimer is in the {\it isosceles} (I) configuration. The Hilbert space of 
$H_{t}$ comprises eight states: a quartet, and two doublets. The quartet ($Q$) 
eigenstates
each one with total spin 3/2, are given in the real space representation by: $\left|
Q,3/2\right\rangle =\left| \uparrow \uparrow \uparrow \right\rangle ,\left|
Q,1/2\right\rangle =\left( \left| \uparrow \uparrow \downarrow \right\rangle
+\left| \uparrow \downarrow \uparrow \right\rangle +\left| \downarrow
\uparrow \uparrow \right\rangle \right) /\sqrt{3},$ while $\left|
Q,-3/2\right\rangle $ and$\left| Q,-1/2\right\rangle $ are obtained from $%
\left| Q,3/2\right\rangle $ and $\left| Q,1/2\right\rangle $ under the
change $\uparrow \longleftrightarrow \downarrow ,$ respectively. It can be
checked that $H_{t}\left| Q,S_{z}\right\rangle =(J_{1}+2J_{2})/4\left|
Q,S_{z}\right\rangle ,$ with $S_{z}=\pm 3/2,\pm 1/2.$ The eigenstates of the
two doublets ($D$ and $D^{\prime }$), each one with total spin 1/2 are: $\left| 
D,1/2\right\rangle
=(\left| \uparrow \uparrow \downarrow \right\rangle -\left| \downarrow
\uparrow \uparrow \right\rangle )/\sqrt{2},$ $\left| D^{\prime
},1/2\right\rangle =(\left| \uparrow \uparrow \downarrow \right\rangle
-2\left| \uparrow \downarrow \uparrow \right\rangle +\left| \downarrow
\uparrow \uparrow \right\rangle )/\sqrt{3},$ with $\left|
D,-1/2\right\rangle $ and $\left| D^{\prime },-1/2\right\rangle $ obtained
again by switching up and down spins. $H_{t}\left| D,S_{z}\right\rangle
=(-3J_{1}/4)\left| D,S_{z}\right\rangle ,$ $H_{t}\left| D^{\prime
},S_{z}\right\rangle =(J_{1}/4-J_{2})\left| D^{\prime },S_{z}\right\rangle ,$
with $S_{z}=\pm 1/2.$ For $J_{1},J_{2}>0,$ as obtained from the SW
transformation, the ground state (GS) of the isolated trimer as a function of $J_{2}$
changes from $\left| D,S_{z}\right\rangle $ to $\left| D^{\prime
},S_{z}\right\rangle $ at $J_{2}=J_{1}$, {\it i. e.} the equilateral case. It is important to note 
that even
for the coupled chain-trimer Hamiltonian, Eq.(5), the total spin of the
trimer remains a good quantum number of the whole system and this is why we can 
characterize the states in the complete system by the total spin of the trimer.

\begin{figure}[tbp]
\includegraphics[width=0.34\textwidth]{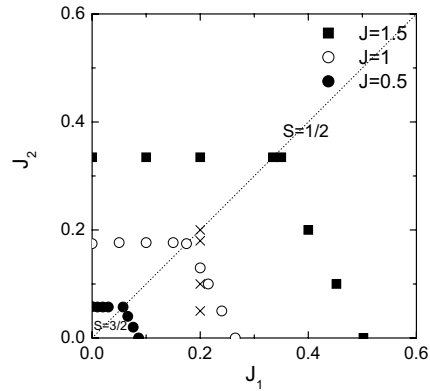}
\vspace{-0.4cm} 
\caption{Boundaries between the strong (large spin) and weak (low spin)  Kondo states 
for three different values of $J$. For each boundary, the large 
(small) $J_1$, $J_2$ values correspond to the low (high) spin values, $S=1/2$ and 
$3/2$ respectively. The crosses mark the positions where the local DOS has been 
calculated (Fig. 4).}
\end{figure}

In Fig. 2 we present the phase diagram of the trimer coupled to the chain, showing
the critical lines
separating both total spin ground states in the trimer (3/2 and 1/2), for different
values of $J$ and $N=13$, the total length of the tight binding chain. The 
phases were obtained by comparing ground state energies calculated using exact 
diagonalization.
For small interaction parameters the system is in the 
high-spin state (below each curve) and for large parameters it is in the 
low-spin 
state (above the curves). The high spin trimer state is increasingly Kondo-stabilized as 
$J$ increases.
The critical boundary has also a different behaviour at the left or right of the 
equilateral line $J_1=J_2$. To the left of this line, the boundary is horizontal, 
{\it  i.e.} for a given value of $J$, the critical value of $J_2$ (defined as 
$J_{2}^{c}$) 
is constant for a large range of $J_1$. In this region the transition occurs between 
the total spin trimer states $\left|Q\right\rangle$ and $\left|D^{\prime}\right\rangle$ 
when increasing $J_2$ for fixed $J_1$ (see case (b) below Eq. (8)). To the right of 
the equilateral line, the boundary is linear and, increasing $J_2$ similarly as 
above, the system evolves from the $\left|Q\right\rangle$ trimer state to the second 
doublet state $\left|D\right\rangle$. For a larger value of $J_2$ (not shown), it 
then turns finally into state $\left|D^{\prime}\right\rangle$ (case (a) explained 
below Eq. 8)).

The scaling behaviour of this phase diagram is shown in Fig. 3, where we present two 
magnitudes which serve as
useful characterizations of the whole system. In Fig. 3(a) we plot the Kondo energy 
$E_{K}({\bf S})\equiv
E_{GS}(N,{\bf S})-\left[ E_{chain}(N)+E_{trimer}({\bf S})\right]$.
Note that by substracting $E_{chain}(N)$ (defined as the
energy of the tight-binding chain with $J=0$) and $%
E_{trimer}({\bf S}),$ what is left is the non-trivial component of the GS
energy related to the Kondo coupling among the magnetic sites of the trimer
and the chain. Accordingly, we denote this energy as $E_{K}({\bf S}),$ with $%
{\bf S}$ representing the total spin of the trimer. As can be seen, the Kondo energy 
for the large spin state is larger in magnitude than for the low spin state.
In Fig. 3(b) we show the scaling behavior of $(J_{2}^{c})$. For values 
of $J_1$ smaller than or equal to the equilateral case, the value of $J_2^c$ 
remains the same (see also Fig. 2), whereas for the opposite case, $J_2^c$ is smaller 
(full dots).

\begin{figure}
\includegraphics[width=0.34\textwidth]{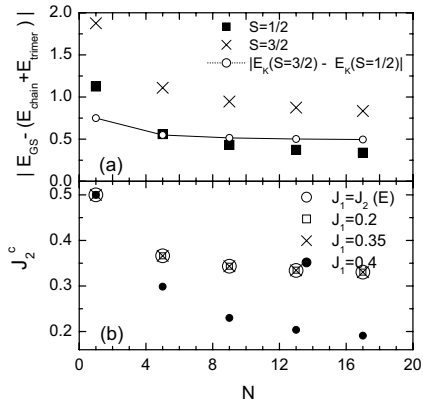}
\vspace{-0.4cm} 
\caption{Scaling behaviour of a) Absolute value of the Kondo energy $E_{K}({\bf S})$ for 
both spin 
states of the trimer and their difference (here  $J_1=0.35$); b) the critical 
$J_2$  $(J_{2}^{c})$ value marking the transition between both spin states, for 
several values 
of the parameters. In both panels the high spin (low spin) state corresponds to  
$J_2<J_{2}^{c}$ ($J_2>J_{2}^{c}$) and $J=1.5$.} 
\end{figure}

In order to understand the results displayed in Figs. 2 and 3 it is quite
instructive to solve the case with one conduction site, $N=1,$ {\it i. e.} the tetrahedron 
(considering one conduction electron).
Noting that $%
E_{chain}(N=1)=0,$ and considering that the trimer could be either in a state with 
total spin 3/2 or 1/2, we have three possible states for $E_{GS}(1,{\bf S}%
)$: 
\begin{eqnarray}
E_{GS}(1,Q) &=&\frac{1}{4}(J_{1}+2J_{2})-\frac{5}{4}J,  \eqnum{4a} \\
E_{GS}(1,D) &=&-\frac{3}{4}J_{1}-\frac{3}{4}J,  \eqnum{4b} \\
E_{GS}(1,D^{\prime }) &=&\frac{1}{4}(J_{1}-4J_{2})-\frac{3}{4}J.  \eqnum{4c}
\end{eqnarray}
The second term which appears on the r.h.s. of Eqs. (6-8), corresponds
either to $E_{K}(3/2)=-5J/4,$ or $E_{K}(1/2)=-3J/4.$ These are just the
binding energies corresponding to the {\it s-d }(Kondo){\it \ }interaction
term of Eq.(5), for the case of a localized magnetic moment with total spin
1/2 $(-3J/4)$ or 3/2 $(-5J/4)$. For $J=1.5,$ $E_{K}(3/2)=-1.875,$ $%
E_{K}(1/2)=-1.125,$ corresponding to the two $N=1$ points in Fig. 3a. As $N$
increases, both $E_{K}(3/2)$ and $E_{K}(1/2)$ evolve smoothly and their difference is nearly 
converged for  $N=17$ (Fig. 3a). From Eqs. (6-8) we can calculate
$J_{2}^{c}$ for $N=1,$ and qualitatively follow its evolution with $N.$
For fixed values of $J$ and $J_{1},$ two scenarios for the GS are possible for 
increasing values 
of $J_{2}$: a) $E_{GS}(Q)\rightarrow
E_{GS}(D)\rightarrow E_{GS}(D^{\prime })$. From $E_{GS}(Q)=E_{GS}(D)$ we
obtain $J_{2}^{c}=J-2J_{1};$ b) $E_{GS}(Q)\rightarrow E_{GS}(D^{\prime }).$
From $E_{GS}(Q)=E_{GS}(D^{\prime })$ we obtain $J_{2}^{c}=J/3.$ The $%
J_{2}^{c}$ corresponding to the cases $J_{1}=J_{2}$, ({\it i.e.} always in the 
equilateral configuration), $J_{1}=0.2$, and 0.35 all fall
in case b) and the three curves fall onto a single curve (Fig. 3b). For $J_{1}=0.4,$
however, case a) is pertinent, and $J_{2}^{c}$ displays a distinct behavior. Both 
kinds of transitions can be seen also in a different behaviour of the boundary lines 
in Fig. 2.

The fact that the total spin of the trimer presents two different values, 
leads to a well defined two-state pattern in the Kondo behaviour of the system. As we showed before 
with the energies, the Kondo energy is stronger for the high spin state. This can be 
clearly seen in 
the local density of states (DOS) at the central site, where it should present a dip when the Kondo 
efect is operative \cite{T1,fano}

In Fig. 4 the local density of states at the central site of
the chain is shown. To obtain the density of states $\rho _{\sigma }(\omega )$ we
use a combined method. In the first place we consider an open finite cluster
of $N$ sites ($N=13$ in our case) which includes the trimer. This is
diagonalized using the exact diagonalization Lanczos technique\cite{lanczos}%
. We then proceed to embed the cluster in an external reservoir of
electrons, which fixes the Fermi level of the system, attaching two
semi-infinite leads to its right and left\cite{ferrari}. This is done by
calculating the one-particle Green function $\hat{G}$ of the whole system
within the chain approximation of a cumulant expansion\cite{cumulant} for
the dressed propagators. This leads to the Dyson equation $\hat{G}=\hat{g}+%
\hat{T}\hat{G}$, where $\hat{g}$ is the cluster Green function obtained by
the Lanczos method. The density of states is obtained from $\hat{G}$.

As a comparison, we have also included in Fig. 4 the LDOS of the isolated 
non-interacting
chain, corresponding to the case $J=0$ in our model. A profound dip appears
in the LDOS of the chain interacting with the trimer, with the LDOS being exactly
zero at the Fermi level of the TB chain $(\omega =0).$ Quite generally, dips
of this type arise each time a system with a continuous spectrum ({\it i.e.}
 the chain), interacts with a system with a discrete spectrum ({\it i.e.}
the trimer)\cite{fano}. While the original discussion of Fano only addressed the case
where the discrete spectrum corresponds to a non-interacting system, his
analysis has been generalized to the case of discrete many-body peaks
interacting with a featureless continuum spectrum\cite{li}. The resulting dips 
in the
DOS are generally termed ``Fano resonances''. Two types of Fano resonances
are observed in Fig. 4. The broad Fano resonance is
associated with the trimer in the state with total spin 3/2 and its width is given by 
$E_{K}(3/2)$.
The narrow Fano resonance is associated with the trimer in the
state with total spin 1/2, and its width is given by $E_{K}(1/2).$ Two
important features should be noticed from Fig. 4: i) the size (width) of the
dip only depends on the total spin of the trimer, and not on the particular
values of $J_{1}$ and $J_{2},$ as long as the trimer total spin remains
invariant. This leads to the bimodal distribution of Fano resonances showed
in Fig. 4; ii) the case $J_{1}=J_{2}=0.2$ corresponds to the trimer in the
equilateral configuration, with total spin 1/2 and a narrow resonance. For a small
reduction in $J_{2}$ $(J_{2}=0.1)$, the trimer is driven to an isosceles 
configuration,
with an increase of its Kondo energy and
the corresponding spectral feature. 

\begin{figure}[tbp]
\includegraphics[width=0.4\textwidth]{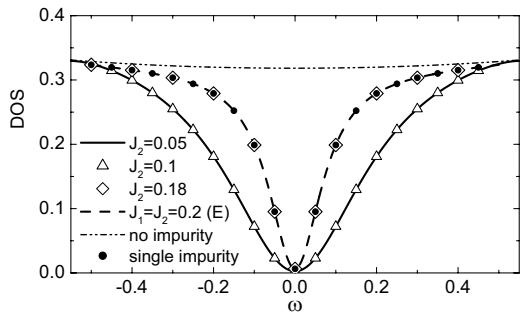}
\vspace{-0.4cm} 
\caption{Local density of states at the central site of the chain for 
$J_1=0.2$, $J=1$ and several values of $J_2$, showing the two-state behaviour of the 
Kondo effect. Also shown is the $J=0$ case (free chain) where no Kondo dip is 
present and the single impurity case. For these parameters $T_K(E)<T_K(I)$.} 
\end{figure}

Recent experimental
results by Jamneala {\it et. al.},\cite{jamneala} investigating the local
behavior of single triangular Cr trimers deposited on the surface of gold show 
that Cr trimers can be reversibly switched between two distintic electronic
states. According to their structural and spectroscopic
characterization, one of the two ``switching states'' corresponds to the ``I''
configuration which, by using the STM {\it dI/dV} spectroscopy, reflects a
low-energy Kondo response (Fano resonance). The second ``switching state''
was characterized as the trimer in the ``E'' configuration, for which no
low-energy feature was discernible. 
Our numerical results, based on a simplified model where the hibridization to a 
single channel is considered, could shed light on these 
experimental results. It captures the essential
physics of this experiment and reproduces its main gross experimental
features: the two-state spectroscopic behavior, and the fact that $%
T_{K}(I)=E_{K}(3/2)>T_{K}(E)=E_{K}(1/2)$, as displayed in Fig. 4. 
For any $T$
such that $T_{K}(I)>T>T_{K}(E)$, the narrow Fano resonance associated to the
``E'' configuration will be lost, as well as the single impurity case, as 
experimentally observed. The ``I''
configuration will exhibit, on the contrary, a discernible Fano
resonance. This behaviour is obtained when all  the antiferromagnetic parameters $J$, 
$J_1$ and $J_2$  are of the same relative order of magnitude\cite{chen}, where the 
magnitude is a fraction of an eV. 
In addition, the isosceles state is obtained from the ``E'' 
configuration, as in the experiment, by slightly reducing the value of $J_2$, {\it i.e.} 
separating atom number 2. 
Our results also agree qualitatively with recent results based on 
Quantum Monte Carlo calculations\cite{savkin}.

Previous calculations on similar models reach contradictory conclusions 
\cite{savkin,aligia,kudasov,lazarovits} finding that the visible Kondo effect 
corresponds to either equilateral or isosceles configurations.
In \cite{kudasov,lazarovits}, calculations based on approximate variational and {\it 
ab initio} methods, restricted
the Hilbert subspace to the one with a Kondo
singlet in the ``trimer+conduction state" (states with energies given by Eqs. 7 and 
8), and missed the ``ferromagnetic''
state of the trimer, driven by the Kondo interaction (state with energy given by 
Eq. 6).
Accordingly, the two-state behavior is not present in their results, and
moreover, both obtain that $T_{K}(E)>T_{K}(I)$, which does not agree with the
experimental interpretation. 

This system is an ideal scenario for the experimental
observation of a quantum critical point between two states with different
quantum numbers and its analysis with finite temperature could lead to
interesting physics\cite{vojta}. 
The rising activity in experimental and theoretical studies of triple quantum 
dots\cite{saraga} is due to the interesting physics that can be 
obtained in these systems. The feasible construction of
a trimer of quantum dots coupled to leads which can be tuned between
different quantum states, could lead to clean and controlable results.
Conductance measurements performed in such a system could detect between distinct 
spin states in the trimer and serve as a readout method.

In conclusion, we have obtained a two-state behaviour in a system consisting
of a spin S=1/2 trimer Kondo-coupled to a non-interacting chain. This
simplified model explains qualitatively the experimental findings of Cr
trimers deposited on the surface of Au, where, at a certain temperature, a
Fano dip was found for the isosceles configuration, but was absent for the
equilateral state, indicating a possible lower Kondo temperature in this latter case,
as obtained here. Moreover, such a system, experimentally realizable as a triangular 
quantum dot device, could be used to measure the total spin state of the timer 
adatom.

The authors acknowledge CONICET for support under grant PIP 0473/98 and
ANPCyT under PICT97 03-00121-02152. \\ \\

\end{document}